\documentclass[11pt,twoside]{article}
\usepackage{./asp2014}

\aspSuppressVolSlug
\resetcounters

\begin{document}

\title{New Parameter Space for Deep Field Radio Continuum Surveys}
\author{Amy J. Barger,$^1$ Kotaro Kohno,$^2$ Eric J. Murphy$^3$, Mark T. Sargent$^4$, and James J. Condon$^3$}
\affil{$^1$University of Wisconsin-Madison, Madison, WI, USA; \email{barger@astro.wisc.edu}}
\affil{$^2$University of Tokyo, Tokyo, Japan; \email{kkohno@ioa.s.u-tokyo.ac.jp}}
\affil{$^3$National Radio Astronomy Observatory, Charlottesville, VA, USA; \email{emurphy@nrao.edu, jcondon@nrao.edu}}
\affil{$^4$University of Sussex, Brighton, UK; \email{Mark.Sargent@sussex.ac.uk}}

\paperauthor{Amy J. Barger}{barger@astro.wisc.edu}{0000-0002-3306-1606}{University of Wisconsin-Madison}{Department of Astronomy}{Madison}{WI}{53706}{USA}
\paperauthor{Eric J. Murphy}{emurphy@nrao.edu}{0000-0001-7089-7325}{National Radio Astronomy Observatory}{}{Charlottesville}{VA}{22903}{USA}
\paperauthor{James J. Condon}{jcondon@nrao.edu}{0000-0003-4724-1939}{National Radio Astronomy Observatory}{}{Charlottesville}{VA}{22903}{USA}
\paperauthor{Kotaro Kohno}{kkohno@ioa.s.u-tokyo.ac.jp}{}{University of Tokyo}{Institute of Astronomy}{Tokyo}{}{181-0015}{Japan}
\paperauthor{Mark T. Sargent}{Mark.Sargent@sussex.ac.uk}{}{University of Sussex}{Astronomy Centre, Department of Physics and Astronomy}{Brighton}{}{BN1 9QH}{UK}

\begin{abstract}
Deep continuum surveys at radio wavelengths are able to cover large areas, yield high angular resolution, and do not suffer from dust extinction, thus providing a robust way to measure the star formation history of the universe.  
However, with the current sensitivities of existing radio telescopes, it remains challenging to detect galaxies that dominate the cosmic star formation history even with extremely long integrations.  
With the ngVLA, a new portion of parameter space will be opened up for radio continuum surveys: deep ($\sim$200\,nJy/bm), large-area ($\sim$1\,deg$^{2}$), sub-arcsecond surveys at high frequencies ($\sim$8\,GHz), where the observed radio emission from high-$z$ galaxies should be dominated by free-free emission, providing a robust measurement for the star formation history of the universe.  
By being able to image the star formation activity, unbiased by dust, for a large, homogeneous sample of galaxies with a wide range of luminosities into the epoch of reionization, such surveys with the ngVLA will be highly complementary to those conducted by {\it JWST}, which will only be sensitive to un-obscured star formation.  
\end{abstract}

\section{Introduction}

The rest-frame UV light density has been successfully mapped to beyond $z\sim10$ using the {\em Hubble Space Telescope\/}.
However, very luminous galaxies emit much of their light at far-infrared (FIR) to millimeter (mm) wavelengths and may be extremely faint in the UV/optical due to large amounts of obscuring dust.  
Thus, long-wavelength observations are needed to determine how much star formation we are missing from UV/optical surveys in these populations of heavily enshrouded systems. 
Indeed, the universe produces comparable amounts of energy in the FIR and optical (Puget et al.\ 1996; Fixsen et al.\ 1998), requiring a detailed understanding of the FIR population at all redshifts to understand how galaxies grow and evolve over cosmic time. 

Uniformly selected samples are key for determining the dusty star formation history, and there are various ways we can try and construct such samples. 
One route is to observe at submm wavelengths, which are unique in that they are equally sensitive to high and low redshifts due to the negative $K$-correction (i.e., with increasing redshift, the fixed submm bandpass samples higher on the Rayleigh-Jeans tail of the blackbody emission produced by the galaxy's dust, scaling as $\nu^2$, which compensates for the inverse square law dimming). Imaging surveys with single-dish submm and mm telescopes 
and cameras have the advantage of being able
to cover large fields and produce uniformly 
selected samples in a single bandpass to a well-established flux limit.
Their main disadvantage is low resolution. 
This is particularly problematic for blank field observations, since the confusion limit\footnote{Confusion refers to the blending of sources and/or the noise 
being dominated by unresolved contributions from fainter sources.} can be reached at fairly high fluxes.
In contrast, interferometric submm and mm imaging surveys enjoy high spatial resolution and sensitivity, but their fields-of-view are small. 

An alternative route is to conduct surveys at radio wavelengths.
The advantages of such surveys are large areas, high spatial resolution, and no extinction. However, at current sensitivities, the positive $K$-correction at these wavelengths results in redshift limitations ($z\lesssim3$) for even the most luminous galaxies.
Differentiating star-forming galaxies from active galactic nuclei (AGNs) at radio wavelengths is also challenging.

Fortunately, the ngVLA is poised to revolutionize radio continuum surveys and make it possible to construct a complete star formation history.
In this chapter, we describe an ngVLA survey that could complement the {\em James Webb Space Telescope (JWST)\/} deep fields by providing high-resolution imaging of star formation, unbiased by dust, for a large, homogeneous sample of galaxies with a wide range of luminosities and star formation rates (SFRs) that probe into the epoch of reionization (i.e., $z\gtrsim6$).

\section{Scientific Importance and Survey Considerations}

Single-dish submm imaging surveys have revealed a significant population of extremely luminous, dusty star-forming galaxies at the highest redshifts 
(e.g., see Casey et al.\ 2014 for a review).  
These galaxies have dramatically changed our understanding of the star formation history of the universe and have challenged cosmological simulations in producing them.  
At some point, the submm selected population will begin to overlap in SFR with the optically selected population.
To obtain large samples of highly obscured star-forming galaxies to the depths necessary to observe
the overlap with the optical population, we can use next generation 
radio continuum surveys that cover large areas, have high spatial resolution, and 
are insensitive to extinction.
In order to determine the optimal frequency continuum survey to be made with the ngVLA, we need to consider the dominant physical mechanisms that contribute to the radio emission \citep[e.g., see this volume, p. ][]{murphy18_ngvla} at high redshifts.

\articlefigure{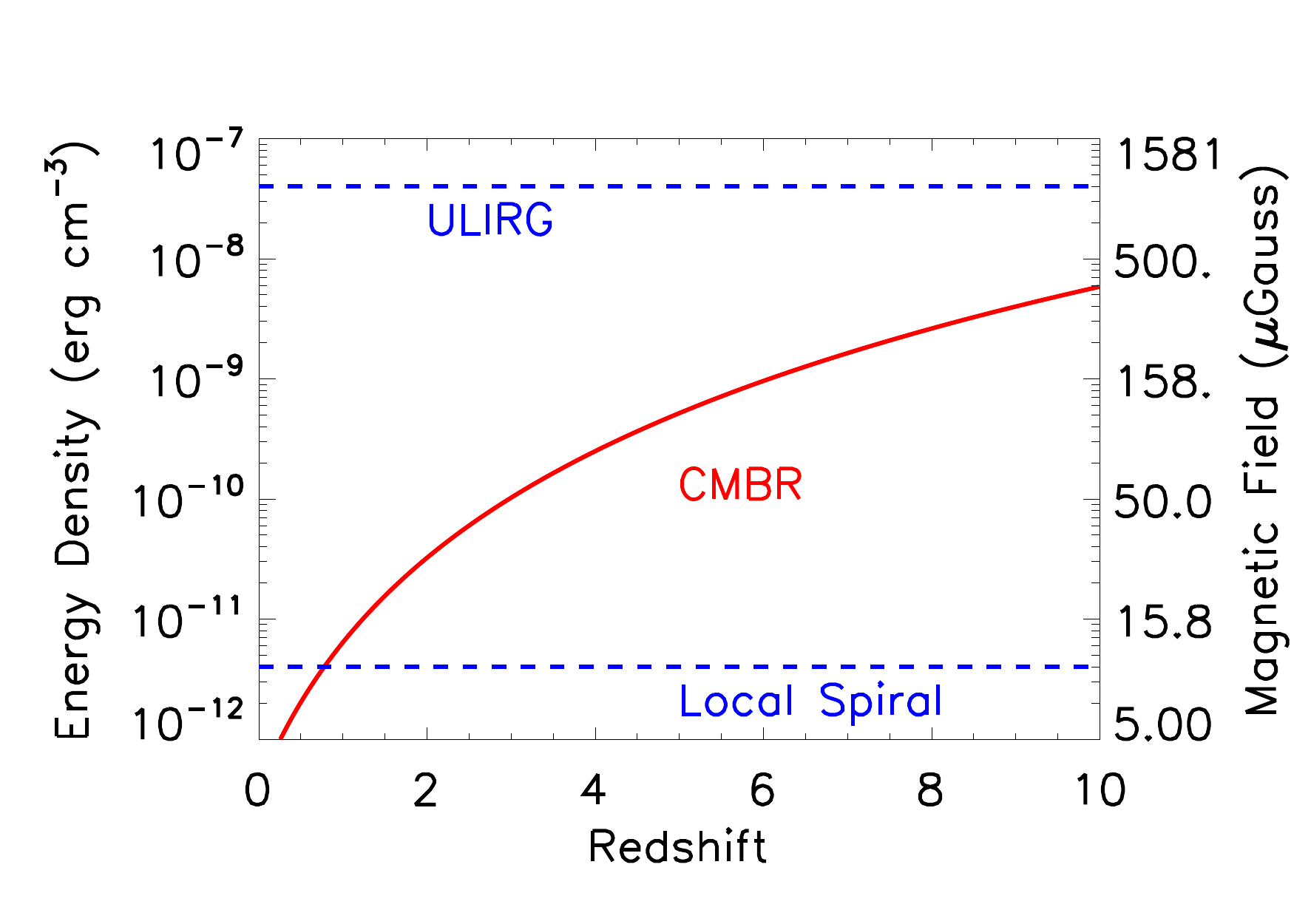}{Fig1}
{Energy density vs. redshift for a local spiral and ULIRG (Carilli 2001; McBride et al.\ 2014), and for the redshift-dependent CMBR.  The corresponding magnetic field strength is shown on the right-hand axis.}

From the well-established FIR-radio correlation (e.g., Helou et al. 1985; Condon 1992), we know that non-thermal synchrotron emission is a valuable tracer of the total amount of star 
formation in galaxies, unbiased by dust. 
Synchrotron emission is dominant in lower frequency radio observations ($\sim1$\,GHz).
The large fields-of-view at these frequencies are an advantage for measuring substantial numbers of very high-redshift galaxies; however,
AGNs are a contaminant in the star-forming galaxy samples. Moreover, there are uncertainties in the conversion from radio flux density to SFR.
The sensitivity of current facilities at these frequencies is very limiting. 
For example, one of the deepest existing 1.4\,GHz VLA images ($5\sigma \approx11\,\mu$Jy of the {\em Chandra\/} Deep Field-North; Owen 2018) is only able to probe luminosities of ultraluminous infrared galaxies or ULIRGs ($L_{IR}\gtrsim10^{12}\,L_\odot; \mathrm{SFR} \gtrsim 125\,M_{\odot}\,\mathrm{yr}^{-1}$) to $z\sim3$ and luminous infrared galaxies or LIRGs ($L_{IR}\gtrsim10^{11}\,L_\odot; \mathrm{SFR} \gtrsim 12\,M_{\odot}\,\mathrm{yr}^{-1}$) to $z\sim 1$ (see Figure~14 in Barger et al.\ 2014).

Inverse Compton (IC) scattering of relativistic electrons off of cosmic microwave 
background radiation (CMBR) photons may also have an impact on synchrotron emission.
When the CMBR radiation energy density exceeds the galaxy magnetic energy 
density, the IC scattering dominates (e.g., Condon 1992; Murphy 2009). 
There is a redshift dependence in the amount of IC scattering that occurs,
since the energy density of the CMBR goes as $\sim (1+z)^4$.
There is also a luminosity dependence, since the synchrotron emission will be highly quenched in moderate SFR galaxies at high redshifts, but possibly never in ULIRGs where the magnetic energy density is higher. 
This effect is illustrated in Figure~1, where we plot energy density versus redshift for a local spiral and 
a ULIRG (from McBride et al.\ 2014 following Carilli 2001) and for the redshift-dependent CMBR. We show the magnetic field strength on the right-hand vertical axis.

\articlefigure{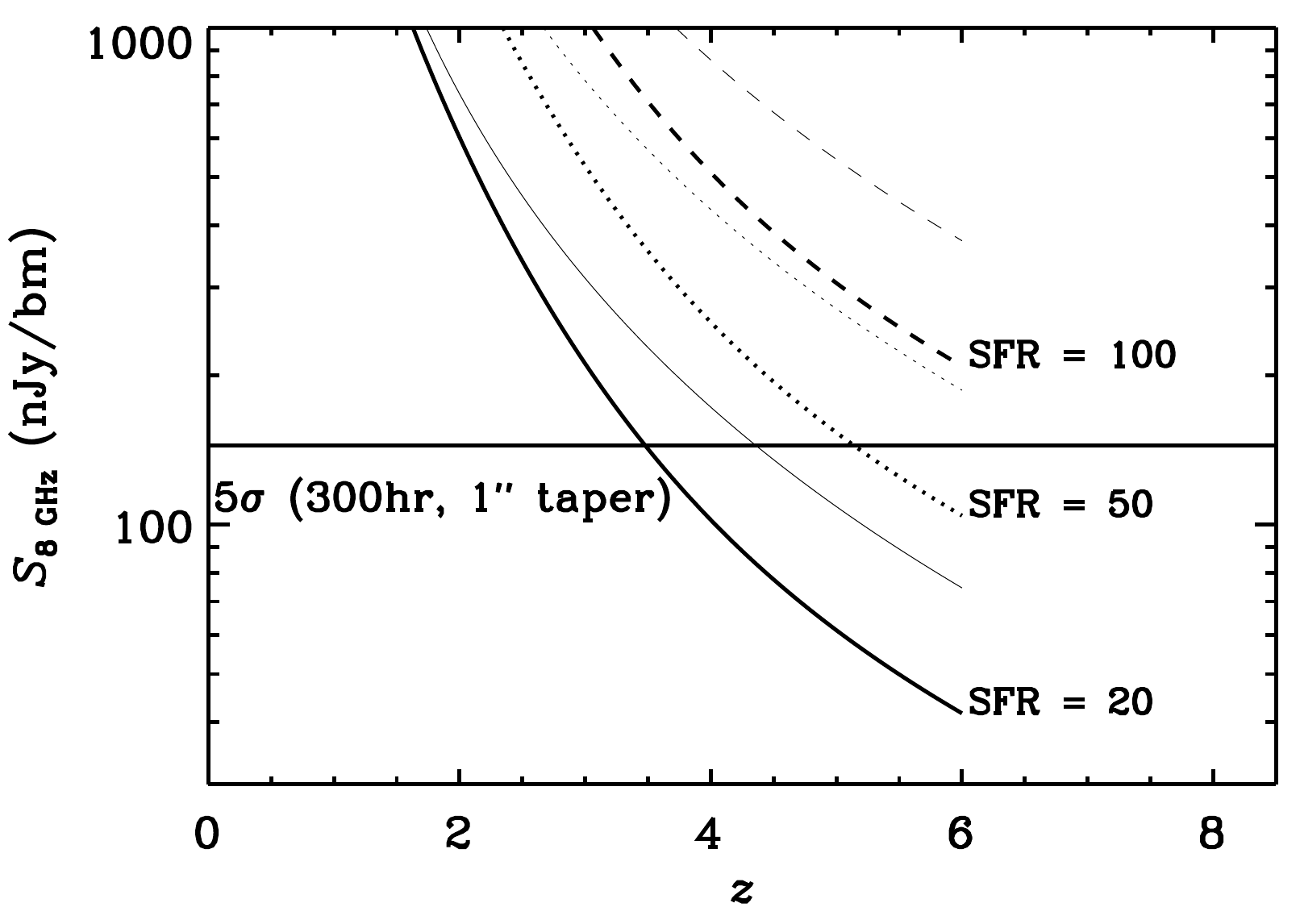}{Fig2}
{Telescope sensitivity in nJy/bm at 8\,GHz plotted against redshift indicating the expected brightness of a 4\,kpc disk galaxy forming stars at a rate of 20, 50, and 100\,$M_{\odot}\,\mathrm{yr}^{-1}$. 
The heavy-weighted lines include estimates for synchrotron dimming due to IC scattering of cosmic ray electrons in galaxies due to the increasing CMBR temperature with redshift, while the corresponding lighter-weighted lines indicate the expected brightnesses in the absence of any synchrotron dimming due to CMB effects. 
Given the current sensitivity specifications, by tapering to a 1\arcsec~synthesized beam, the ngVLA will have enough brightness temperature sensitivity to detect galaxies forming  stars at a rate of $>50\,M_{\odot}\,\mathrm{yr}^{-1}$ into the Epoch of Reionization.  
Using the existing VLA, the detection of such a galaxy would take $\gtrsim$10,000\,hr!}

Consequently, this means that the emission detected from high-redshift galaxies with moderate SFRs should be dominated by free-free emission, modulo the possible contribution from anomalous dust emission \citep[e.g.,][]{ejm10,ejm18,planck15}, which may arise from spinning dust grains  (e.g., Draine \& Lazarian 1998).  
This in turns results in a rather ideal situation for attempting to cleanly measure star formation rates form high-redshifts systems.  
Free-free emission, which is dominant at rest-frame tens of GHz \citep{condon92}, is directly proportional to the production rate of ionizing photons by young, massive stars, making extremely clean for measuring accurate SFRs (see, e.g., Murphy et al.\ 2015).  

\section{Anticipated Results and Astronomical Impact}

The ngVLA will open up completely new parameter space for radio continuum surveys, allowing for the extremely deep, large-area mappings to be made at much higher frequencies than traditional surveys at $\sim$1\,GHz.  
For instance, by covering the anticipated maximum size of potential {\em JWST\/} deep fields ($\lesssim$1\,deg$^2$), which are the most heavily studied regions on the sky, to a sensitivity of 0.2~$\mu$Jy per beam at  $\approx$8\,GHz ($1\arcsec$ resolution), highly robust, dust-unbiased SFRs of the galaxies in these fields can be obtained. 
In particular, such a survey will allow the detection of many thousands of $L^\ast$ (i.e., $\sim 2\times 10^{12}\,L_\odot; \mathrm{SFR} \gtrsim 25\,M_{\odot}\,\mathrm{yr}^{-1}$) galaxies at $z\sim 2$ to 4\, as well as galaxies with $L_{IR}\gtrsim5\times 10^{12}\,L_\odot; \mathrm{SFR} \gtrsim 625\,M_{\odot}\,\mathrm{yr}^{-1}$ out to $z\sim 8$, where we currently have little constraints on the potential numbers of such luminous galaxies.  
Furthermore, using the same data, but applying a different imaging weighing scheme to create finer resolution maps (i.e., 0\farcs1, or even higher for brighter systems), one can obtain information on morphologies of these systems, revealing highly obscured regions that could be missed in rest-frame optical/UV light form {\em JWST} imaging.  

For an ultra-deep integration (i.e., $\approx$300\,hr; $\mathrm{rms}  \approx 30$\,nJy/bm) the ngVLA should be able to detect a galaxy at the $5\sigma$ level with a SFR of $\sim20\,M_{\odot}\,\mathrm{yr}^{-1}$ out to $z\sim3$, and a galaxy with a SFR of $\sim100\,M_{\odot}\,\mathrm{yr}^{-1}$ to $z\sim7$.
This is illustrated in Figure \ref{Fig2}, which shows the observed 8\,GHz brightness of 4\,kpc diameter disk galaxies with a range of SFRs. 
The 5\,$\sigma$ 8\,GHz rms of the ngVLA tapered to a 1\arcsec~synthesized beam is based on the updated sensitivities given in \citep{memo17}\footnote{\url{http://ngvla.nrao.edu/page/refdesign}}. 
Consequently, given the current sensitivity specifications, by tapering to a 1\arcsec~synthesized beam the ngVLA will have enough brightness temperature sensitivity to detect galaxies forming  stars at a rate of $>50\,M_{\odot}\,\mathrm{yr}^{-1}$ into the epoch of reionization.  
Using the existing VLA, the detection of such a galaxy would take $\sim$10,000\,hr.

By combining the data for a large sample of 8\,GHz selected galaxies with observations at other frequencies from 1 to 50\,GHz that could also be made with the ngVLA, one could map the radio spectral energy distributions (SEDs) of the galaxies, which would make it possible to robustly account for the different thermal and non-thermal energetics by separating the free-free and potential synchrotron contributions.  
The spectral indices could also be used to separate AGN form star-forming galaxies in surveys.  
This is critical for the construction of the total star formation history from radio data. 
Another interesting outcome of mapping the radio SEDs for a number of galaxies of a given luminosity is to determine at which redshift the spectrum becomes purely free-free emission, resulting in a rough measurement for the magnetic field strength of galaxies at that luminosity by equating it to the CMBR energy density at that redshift (see Figure~\ref{Fig1}). 
And, while not the focus of this chapter, it is worth pointing out that with such a large frequency range, we could make measurements of low-$J$ CO lines to obtain redshifts for  many of the galaxies---particularly the more dusty galaxies where it is very difficult to measure redshifts from optical and near-infrared  spectroscopy---and to provide information on the gas masses of the  galaxies.




\section{Summary}
Single-dish submm images have low resolution, and blank field observations hit the 
confusion limit at relatively bright fluxes ($\sim2$\,mJy).  Interferometric submm/mm images 
have high spatial resolution and sensitivity but suffer from small fields-of-view.  The
deepest existing 
continuum surveys with the VLA are limited to finding complete samples of ULIRGs to $z\sim3$ 
and LIRGs to $z\sim1$ (e.g., see Figure~14 of Barger et al.\ 2014),
while Milky Way luminosities can only be detected in the local region
($z\ll 0.5$). The unprecedented sensitivity of continuum 
imaging surveys with the ngVLA will push the boundaries of galaxy evolution studies 
considerably (see Figure~2). Moreover, with the ngVLA's broad high-frequency coverage, 
we will be able to estimate the fraction of light being produced by free-free versus 
synchrotron as a function of frequency.  
The ngVLA observations will add enormous value to the {\it JWST} deep fields by providing high-resolution, dust-free images of the galaxies to study the morphologies of the star formation.  
They will also provide complementary observations to SKA1, allowing a better understanding of the radio production mechanisms.   





\end{document}